\documentclass[cameraready]{Interspeech}

\title{How Bilingual Are SSL Speech Models? Cross-Lingual Probing of Articulatory Encoding with Finnish and Russian EMA}

\author[affiliation={1}, orcid=0009-0009-9778-3104]{Ailín}{Pollio San Pedro}
\author[affiliation={2}, orcid=0000-0002-4371-7322, ]{Tomi}{Kinnunen}
\author[affiliation={2}, orcid=0000-0001-8634-5947]{Alexandre}{Nikolaev}
\author[affiliation={2}, orcid=0009-0002-6627-2706]{Ruchi}{Pandey}

\address{
    $^1$ University Grenoble Alpes, CNRS, Grenoble INP, GIPSA-lab, France \\
    $^2$ University of Eastern Finland, Computational Speech Group, Finland 
}

\email{$^1$ailin.pollio-san-pedro@cnrs.fr, $^2$firstname.lastname@uef.fi}


\usepackage{comment}

\usepackage{tikz}
\newcommand{\expbadge}[1]{%
\tikz[baseline=(char.base)]{
  \node[
    circle,
    draw,
    font=\bfseries,
    inner sep=1.0pt
  ] (char) {#1};
}}

\begin{document}

\maketitle

\begin{abstract}
    SSL speech models capture rich phonetic, prosodic, and acoustic patterns from raw audio, yet how they encode articulatory information across diverse languages remains unclear. Using EMA data from bilingual Finnish–Russian speakers, we evaluate cross-lingual correlations between SSL latent representations and articulatory movements. Models achieve strong prediction performance (Pearson $r$ up to 0.68) even with $\sim$ 5 minutes of training data, with multilingual models outperforming monolingual ones. Intermediate layers encode articulatory features most effectively, and tongue movements are more predictable than lip movements. We also assess the impact of task type (read versus spontaneous speech) and language proficiency, finding higher accuracy for structured tasks and strong generalization across proficiency levels. These results enhance the interpretability of SSL models and show their potential for speech-technology applications.
\end{abstract}
\section{Introduction}
\label{sec:intro}
Self-supervised learning (SSL) models such as wav2vec 2.0~\cite{Baevski2020}, HuBERT \cite{Hsu2021}, and WavLM \cite{Chen2022} have become foundational in modern speech research. Trained on large-scale unlabeled audio, they learn structured representations that transfer effectively to diverse downstream tasks \cite{Yang2021SUPERB}. However, the precise nature of these representations remains only partially understood. Probing studies have revealed that SSL features encode phonetic categories~\cite{Ji2022, Martin2023, Venkateswaran2025}, speaker characteristics~\cite{chiu2025largescaleprobinganalysisspeakerspecific}, and suprasegmental structure (e.g. prosody and tone)~\cite{delafuente24_interspeech},
yet these models are often treated as opaque feature extractors. 

Recent work has investigated whether, and how, SSL representations capture the physical dynamics of speech production. Using electromagnetic articulography (EMA) data and linear probing, 
the authors in \cite{Cho2023} found that SSL representations correlate strongly with articulatory trajectories ($r>0.8$), using as little as five minutes of paired data. 
Another study \cite{Udupa2023} confirmed that SSL features match or exceed traditional MFCCs for acoustic-to-articulatory inversion (AAI), and 
\cite{Wu2023} further demonstrated that SSL representations generalize to unseen speakers more effectively than handcrafted features. Layer-wise analyses further indicate that articulatory information concentrates at intermediate transformer layers, while deeper layers shift toward more abstract linguistic content \cite{Pasad2021}. Intermediate layers also offer the best balance between speaker-invariant encoding and articulatory precision, a property critical for generalization across speakers and speaking styles~\cite{Wu2023}.

The research has further extended to test whether articulatory structure is recoverable when pre-training and probing languages differ. 
The authors in \cite{Cho2024} probed SSL models pre-trained on several languages using EMA data from English, Mandarin, and Italian speakers, reporting consistently high articulatory inversion correlations across pre-training languages. Predicted trajectories could be aligned across speakers via affine transformations, suggesting a shared articulatory subspace with largely geometric inter-speaker differences \cite{Cho2024}. 
Another study \cite{Hao2024} further showed that SSL models retain strong AAI performance in a zero-shot English-to-Dutch transfer setting, particularly when using intermediate encoder layers. 

These results support the hypothesis that SSL representations capture partially language-agnostic articulatory features, although cross-lingual performance may drop when the target language’s phonotactics or prosody differs from the pretraining data. However, languages that are typologically distant from those studied so far remain largely unexplored. 
Our study extends prior SSL–EMA probing beyond monolingual English settings by investigating articulatory encoding in bilingual Finnish–Russian speakers using the recently collected FROST-EMA corpus \cite{Hopponen2025}. 
Finnish, with its vowel harmony and phonemic quantity contrasts, and Russian, with its pervasive palatalization, present a meaningful testbed for claims of language-general articulatory encoding. We systematically evaluate cross-lingual generalization, layer-wise and sensor-wise distribution of articulatory information, and the influence of task structure and language proficiency on prediction accuracy. 
The contributions of this paper are threefold:
\begin{itemize}
    \item We provide the first systematic SSL–EMA probing analysis for Finnish and Russian, two typologically distinct and previously unexamined languages in this context.
    \item We quantify the robustness of articulatory representations across native (L1), second-language (L2), and accent-imitated speech, presenting the first probing analysis across these bilingual conditions.
    \item Unlike prior work focusing on controlled read speech, we analyze layer-wise and sensor-wise encoding across controlled and spontaneous tasks, using different models.
\end{itemize}

\section{FROST-EMA Dataset}
\label{subsec:dataset}
The articulatory dataset used is the recently collected, sizeable FROST-EMA corpus~\cite{Hopponen2025}. 
It consists of parallel acoustic and EMA recordings from 18 bilingual speakers (8 female, 10 male), covering Finnish and Russian as first and second languages, along with imitated foreign-accented speech. The 
age range of the subjects is 21 to 49 years ($M\!=\!30.9$, $SD\!= \!9.7$), 
all speakers being bilingual to varying degrees. Self-reported proficiency in the second language ranged from basic to native-like, and 11 participants reported using their L2 in daily life. 

For the present study, we used data from five active articulators: tongue tip (TT), tongue anteo-dorsum (TB), tongue dorsum (TD), upper lip (UL), and lower lip (LL). For each articulator, the X (back–front) and Z (up–down) coordinates were extracted, yielding 10 continuous articulatory features at a rate of 1250 Hz. The Y axis (lateral motion) is excluded by default due to its limited acoustic relevance and potential interpretational ambiguity (for reference purposes, however, we do include it in selected probing analyses).
%
%
%
The corpus consists of three tasks:
\begin{enumerate}
    \item \textbf{Read speech:} The \emph{North Wind and the Sun} passage.
    \item \textbf{Sentence-level elicitation:} Reading short carrier sentences containing target words - (Finnish: “Sano \texttt{[target]} uudelleen”; Russian: “Povtori \texttt{[target]} snova”, both meaning “Say/Repeat \texttt{[target]} again”), comprising 71 stimuli in Finnish and 75 in Russian.
    \item \textbf{Spontaneous speech:} Comic-strip narration task to elicit natural speech \cite{Gagarina2019}.
\end{enumerate}
Importantly, each task was further completed under three distinct language conditions: (A) in the subject's native language (L1), (B) in their L1 while imitating a foreign accent corresponding to their L2, and (C) in their second language (L2). This $3\!\times\!3$ design yields nine distinct 
conditions per speaker. 
Tasks were selected to span controlled and spontaneous speech, allowing analysis of how increasing linguistic and prosodic variability affects articulatory predictability. Recordings were conducted in a professional-grade soundproof booth using a high-quality shotgun microphone and a state-of-the-art EMA device.
\section{Experimental design}
\label{sec:expdesign}
Speech recordings were passed through each pre-trained SSL speech model, which served as a feature extractor. For each model, we extracted hidden representations from all Transformer layers. All encoders used in this study have 24 layers with 1024-dimensional hidden representations. This resulted in a set of layer-wise feature representations per recording, enabling systematic probing across network depth. A linear regression framework was then used to predict EMA trajectories from the SSL representations, facilitating a fine-grained analysis of articulatory encoding as a function of model layer, language condition, and speaker profile.

\subsection{Preprocessing}
EMA positional files (\texttt{.pos}) were normalized and temporally matched to the SSL feature frame rate. To suppress arbitrary offsets due to coil placement, inter-speaker scale differences, and channel-specific amplitude ranges, we z-normalized each sensor-axis trajectory independently within each recording 
using $\tilde{s}_i = (s_i - \mu_s)/\sigma_s$,
where $s_i$ denotes the $i$-th sample of a given channel within the recording, and $\mu_s$ and $\sigma_s$ are the mean and standard deviation computed over all time frames of that channel in the corresponding utterance. Before decimation to 50\,Hz---the frame rate of the SSL feature extractor---we applied a Butterworth low-pass filter 
to mitigate aliasing 
and to preserve the main articulatory dynamics in the $< 20$\,Hz range. 
\subsection{SSL feature extraction}
Self-supervised encoders learn latent speech representations without 
the need for data labels~\cite{robin2020sslspeech}. For each input frame, we extracted a 1024-dimensional vector from every transformer layer of each model in the Wav2Vec\,2.0 family (including multilingual variants). These latent vectors capture statistically induced regularities in the utterances that have been shown to correlate with articulatory structure~\cite{liu2022sslreview}. For an utterance with $T$ frames (where $T$ depends on the 
utterance duration), the SSL feature sequence extracted by the $\ell^\text{th}$ layer is
$\, (\mathbf{x}^{(\ell)}_t) \in \mathbb{R}^{1024 \times T}$, 
where
$\ell\!=\!1\ldots\!24$ and $t\!=\!1\!\ldots\!T \,$.

\subsection{Articulatory score (linear probing)}
To quantify how well layer-wise SSL representations encode articulation, we trained a \emph{linear probe}~\cite{AlainBengio2016Probing} to predict 10 EMA trajectories (the two X/Z coordinates $\times$ five sensors) from the latent features. 
%
%
%
In particular, the probe is given by
\begin{align}
  \hat{y}_{t}^{(k)} = \mathbf{w}^{(k)\top} \mathbf{x}_t + b^{(k)},
  \label{equation:eq4}
\end{align}
\noindent where $\mathbf{x}_t \in \mathbb{R}^{1024}$ is the SSL representation at frame $t$, 
$\mathbf{w}^{(k)} \in \mathbb{R}^{1024}$ and $b^{(k)} \in \mathbb{R}$ are the regression weights and bias for EMA channel $k$, and $\hat{y}_{t}^{(k)}$ is the predicted value for that channel.

Data from all recordings of a given speaker were concatenated at the frame-level. The resulting dataset was randomly shuffled and split into 80\%/20\% train/test sets using a fixed random seed for reproducibility. 
Performance was evaluated using the Pearson correlation coefficient between the predicted trajectory $\hat{y}_{t}$ and the reference EMA trajectory $y_{t}$ over $T$ time frames.
The correlation coefficient, indicated by $r_{s,d}$, was computed independently for each EMA dimension $d$ and speaker $s$ on the held-out test frames. Following~\cite{Cho2023},
the \emph{articulatory score} 
was then computed as the average correlation across all EMA dimensions ($D=10$) and speakers ($S$):
\begin{align}
\text{Articulatory Score} =
\frac{1}{S D} \sum_{s=1}^{S} \sum_{d=1}^{D} r_{s,d}.
\label{eq:articulatory_score_probe}
\end{align}

\subsection{Experimental setups}
We conducted five complementary experiments to analyze model-, layer-, data-size-, and generalization effects 
(Table~\ref{tab:exp-grid}).

\noindent\textbf{Comparison of SSL models (E1)} addresses how multilingual SSL pretraining and/or language-specific fine-tuning impacts 
the articulatory scores. We include five models that differ in their ``familiarity'' with Finnish and Russian (Wav2Vec\,2.0 Large\footnote{https://huggingface.co/facebook/wav2vec2-large}; MMS-300m\footnote{https://huggingface.co/facebook/mms-300m}; XLSR-53\footnote{https://huggingface.co/facebook/wav2vec2-large-xlsr-53}; XLSR-53 fine-tuned on Russian\footnote{https://huggingface.co/jonatasgrosman/wav2vec2-large-xlsr-53-russian}; and XLS-R fine-tuned on Finnish\footnote{https://huggingface.co/Finnish-NLP/wav2vec2-xlsr-300m-finnish-lm}). For each model, we report a single score obtained by averaging the articulatory scores across EMA dimensions, speakers, and layers.

\noindent\textbf{Sensor--layer profiles (E2)} addresses 
layer specificity of the articulatory scores, and its dependency on the articulator. Using Wav2Vec\,2.0 Large, MMS-300m, and XLSR-53, we compute layer-wise articulatory scores for each EMA dimension across the 24 transformer layers, yielding sensor--layer profiles that localize peaks and drops in articulatory predictability.

\noindent\textbf{Training-size sensitivity (E3)} addresses how data-efficient the linear probe is, i.e., how much speaker-specific EMA data is needed before performance saturates. Using Wav2Vec\,2.0 Large, we vary the probe training duration per speaker (20\,s, 30\,s, 1\,min, 5\,min, 10\,min, up to 20\,min when available) using an otherwise equivalent evaluation protocol. We report articulatory scores as a function of training duration.

\noindent\textbf{Speaker generalization (LOSO; E4)} addresses how well probing performance transfers to unseen speakers. We run leave-one-speaker-out (LOSO)~\cite{hastie2009elements} cross-validation with MMS-300m, training on $S\!-\!1$ speakers and testing on the held-out speaker in each fold. We report per-speaker, per-dimension Pearson correlations to characterize generalization and inter-speaker variability.

\noindent\textbf{Task and proficiency effects (grouped LOSO; E5)}  isolates the effects of linguistic structure (language and accent) versus production variability (task type) on articulatory predictability. We repeat LOSO within Finnish and within Russian cohorts while matching the three tasks 
and the three language conditions 
(explained in Section~\ref{subsec:dataset}).
We report per-speaker, per-dimension correlations stratified by task and condition to quantify how input variability and L2 proficiency modulate predictability.

\begin{table}[t]
\caption{Overview of experimental configurations. LOSO = Leave-One-Speaker-Out.}
\label{tab:exp-grid}
\centering
\footnotesize

\begin{tabular}{l p{0.8\columnwidth}}
\toprule
\textbf{ID} & \textbf{Experimental configuration and evaluation} \\
\midrule

\expbadge{E1} 
 & \textbf{Cross-model comparison}. 
18 speakers (combined). 
Models: WV2-L, MMS-300m, XLSR-53, RU-FT, FI-FT. 
Output: Mean articulatory score across layers and EMA dimensions. \\

\midrule

\expbadge{E2}
& \textbf{Sensor--layer mapping}. 
18 speakers (combined). 
Models: WV2-L, MMS-300m, XLSR-53. 
Output: Pearson $r$ per EMA dimension and layer (1--24). \\

\midrule

\expbadge{E3} 
& \textbf{Training-size sensitivity}. 
18 speakers (combined). 
Model: WV2-L. 
Output: Pearson $r$ vs.\ probe training duration (20\,s--20\,min). \\

\midrule

\expbadge{E4} 
& \textbf{Speaker generalization (LOSO)}. 
$S\!-\!1$ train / 1 test. 
Model: MMS-300m. 
Output: Pearson $r$ per speaker and EMA dimension. \\

\midrule

\expbadge{E5} 
& \textbf{Task \& proficiency effects}. 
LOSO within Finnish/Russian. 
Model: MMS-300m. 
Output: Pearson $r$ per speaker, EMA dimension, task, and condition (L1, L1+accent, L2). \\
\bottomrule
\end{tabular}
\vspace{-0.3cm}
\end{table}

A \emph{per-speaker} probing setup was adopted in Experiments 1-3 to mitigate the effects of known inter-speaker EMA variability arising from sensor placement differences and occasional reattachment events during data collection. 
\section{Results}
\label{sec:results}
This section reports the results for the five experiments summarized in Table~\ref{tab:exp-grid}. 
Unless otherwise noted, scores are Pearson correlations averaged over EMA channels.

\begin{figure*}[t]
    \centering
    \includegraphics[width=0.9\textwidth]{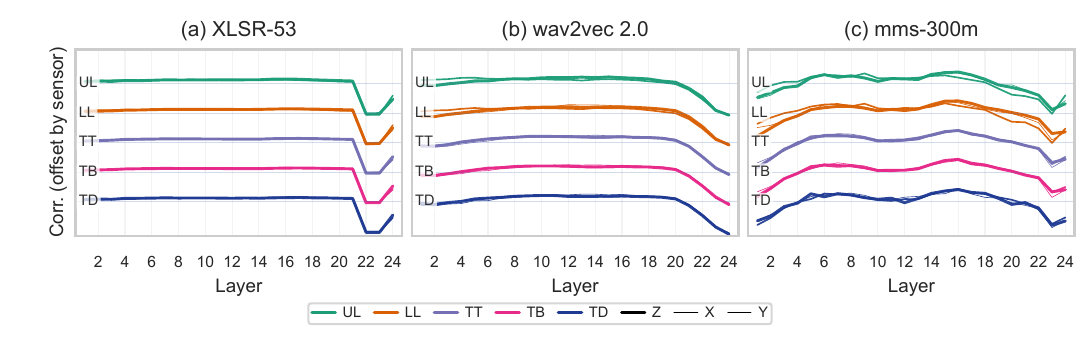}
    \caption{Layer-wise articulatory prediction performance (average Pearson correlation \(r \)) across encoder depth for (a) XLSR-53, (b) wav2vec~2.0, and (c) MMS-300m. Curves are vertically offset by sensor for visual clarity. While distinct correlation trajectories are observed across spatial axes, the overall layer-wise profiles remain highly similar across articulators and coordinate axes, suggesting consistent probing dynamics throughout the network.}
    \label{fig:sensor_evolution}
    \vspace{-0.2cm}
\end{figure*}

\noindent\textbf{E1: Cross-model comparison.} Table~\ref{tab:model-scores} summarizes the average articulatory scores over all speakers, sensors, and layers. MMS-300m and the language fine-tuned XLS variants 
achieve the highest articulatory scores ($r\!\approx\!0.69$), while XLSR-53 produces substantially lower scores ($r\!=\!0.62$). Wav2Vec\,2.0 Large (English) reaches $r\!=\!0.641$. These findings align with the broader coverage of the two languages during MMS pretraining and fine-tuned variants, and contrast with the reduced coverage and late-layer degradation observed for XLSR-53. 
As shown in Fig.~\ref{fig:sensor_evolution}(a),  XLSR-53 exhibits a sharp correlation drop at layers 22–23, unlike the smoother late-layer trends in other encoders. 

\begin{table}[t]
\centering
\caption{Articulatory prediction performance (Pearson $r$) across SSL models.}
\label{tab:model-scores}
\begin{tabular}{l c}
\hline
\textbf{Model} & \textbf{Mean $r$} \\
\hline
Wav2Vec\,2.0 Large (EN) & 0.641 \\
MMS-300m (multilingual) & 0.689 \\
XLSR-53 (multilingual) & 0.620 \\
XLSR-53 (RU fine-tuned) & 0.689 \\
XLS-R (FI fine-tuned) & 0.686 \\
\hline
\end{tabular}
\vspace{-0.3cm}
\end{table}

\begin{figure*}[t]
    \centering
    \includegraphics[width=1.0\textwidth]{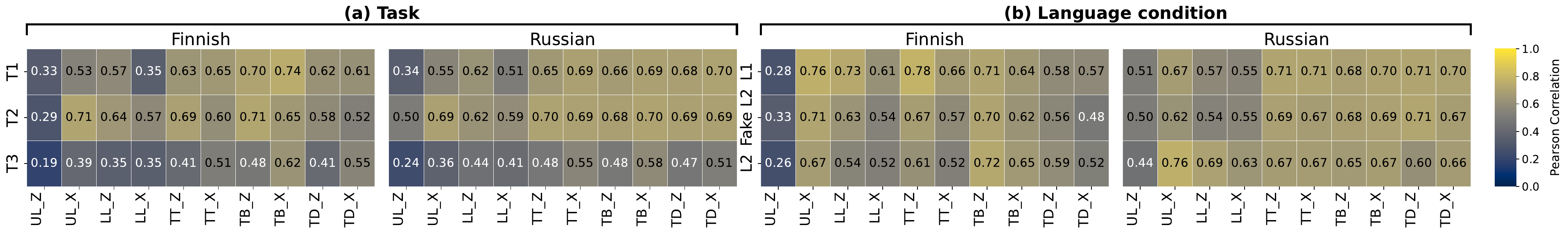}
    \caption{Layer-wise articulatory prediction performance (Pearson correlation $r$) for MMS-300m under a LOSO evaluation. (a) Task comparison across Finnish and Russian speakers, showing higher correlations for controlled reading tasks (T1-T2) than for spontaneous picture description (T3). (b) Language-condition comparison (L1, L2, and simulated accent conditions.)}
    \label{fig:LOSO}
    \vspace{-0.4cm}
\end{figure*}




\noindent\textbf{E2: Layer--sensor profiles.} Figures~\ref{fig:sensor_evolution}(b) and ~\ref{fig:sensor_evolution}(c) show  that articulatory prediction performance varies systematically across encoder depth. For both wav2vec~2.0 and MMS-300m, scores increase toward the middle layers, where they peak, and then decrease in the highest layers, with a sharper drop observed for wav2vec~2.0. While XLSR-53 stays relatively stable up to layer 21, drops sharply at layers 22--23 with a small recovery at layer 24. Across models, X/Z channels are consistently easier to predict than Y, with TB\_Z among the strongest channels and UL\_Z among the weakest.

\begin{figure}[t]
    \centering
    \includegraphics[width=0.8\linewidth]{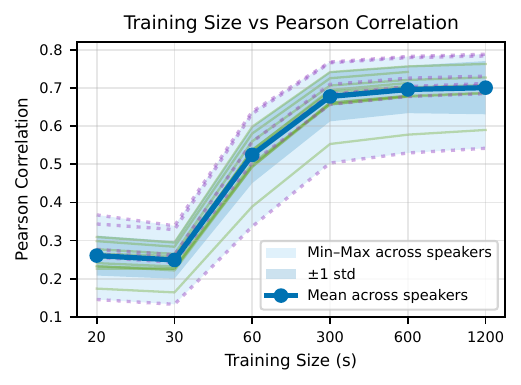}
    \caption{Performance increases with training size and stabilizes around 300 seconds. The solid green lines correspond to Finnish speakers, while the dashed purple lines indicate Russian speakers.}
    \label{fig:training_size}
    \vspace{-0.4cm}
\end{figure}

\noindent\textbf{E3: Training-size sensitivity.} As revealed in Figure \ref{fig:training_size}, 
articulatory scores increase steeply as training duration grows from seconds to a few minutes. This indicates high sample efficiency of SSL representations for articulatory prediction. The scores largely saturate after approximately five minutes of paired data. 

\noindent\textbf{E4: Speaker generalization (LOSO).} LOSO evaluation with MMS-300m indicates robust generalization to unseen speakers, with best-case per-speaker correlations reaching \(r\!\approx\!0.78\). Tongue sensors (TT/TB) generalize most consistently, whereas UL\_Z remains systematically challenging. Variability across speakers is consistent with known sensor-quality differences during data collection.

\noindent\textbf{E5: Task and language effects (LOSO within-group).} As  observed in Fig.~\ref{fig:LOSO}, controlled reading tasks yield higher correlations than spontaneous picture description, suggesting that reduced linguistic variability benefits articulatory probing. Peak channels reach \(r \approx 0.70\)--\(0.74\) in controlled conditions, while picture description drops to roughly \(r \approx 0.58\)--\(0.62\). Notably, L2 speech can match L1 accuracy in several channels (up to \(r \approx 0.76\)), pointing to accent-robust articulatory encoding in MMS-300m, alongside persistent channel-specific limitations (e.g., UL\_Z).

\section{Discussion}

Drawing upon the findings presented above, we have the following four main findings.

\noindent\textbf{FINDING 1: Intermediate layers consistently yield the strongest articulatory alignment.}
Results support the view that SSL encoders capture physically meaningful production dynamics: linear probes reached moderate-to-strong correlations without task-specific fine-tuning of the backbone, and even when target languages were not present during pretraining. Peak performance arose at intermediate layers, consistent with the hypothesis of a hierarchy in which lower/intermediate layers concentrate phonetic–articulatory cues while upper layers drift towards more abstract content \cite{Cho2023,Pasad2021}.

\noindent\textbf{FINDING 2: Task structure modulates articulatory predictability.}
Controlled reading and sentence tasks were found to yield higher correlations than spontaneous picture description, likely because lexical/syntactic variability and heterogeneous utterance lengths in spontaneous speech reduce alignment between acoustics and sensor kinematics. Notably, most prior studies evaluated models on controlled read speech \cite{Cho2023,Wu2023,Hao2024}, leaving the impact of task structure underexplored. This effect was visible across languages and sensors, and was most pronounced for vertical upper-lip movement (UL\_Z), which remained hardest to predict in our experiments.

\noindent\textbf{FINDING 3: Tongue articulators are more reliably encoded than lips.} Tongue trajectories (e.g., TB\_X, TT\_Z) were consistently more predictable than lips in our experiments, aligned with prior probing analyses showing articulator-specific variability in linear decodability from SSL representations \cite{Cho2023}. Horizontal upper-lip movement (UL\_X) could attain competitive correlations in some language/condition cells, indicating that labial gestures remain reliably encoded when task demands constrain segmental inventory and coarticulation.

\noindent\textbf{FINDING 4: Multilingual pretraining improves cross-lingual robustness.}
Grouping by language generally increased scores relative to task-only groupings, suggesting that phonemic inventory, prosodic structure, and segmental patterns modulate how SSL features align with articulation. Multilingual pretraining (e.g., MMS-300m, XLSR-53) improved robustness relative to monolingual encoders, consistent with the hypothesis that broader acoustic–phonetic coverage helps recover language-agnostic articulatory regularities \cite{Pratap2024,Baevski2020}.

\section{Conclusion}

Our case study on SSL model probing using EMA position data of bilingual speakers (Finnish and Russian) contributes to the growing body of evidence of the presence of articulatory dynamics encoding by these models. Overall, the relatively strong correlations (up to \(r \!\approx\!0.78\) in LOSO speaker generalization (MMS-300m), and mean \(r  \!\approx\! 0.69\) in within-speaker probing) reinforce that SSL speech representations encode language-general articulatory structure. Even without exposure to Finnish or Russian during pretraining, monolingual encoders allowed linear recovery of EMA trajectories, while multilingual models improved cross-lingual robustness. The consistent emergence of peak performance at intermediate layers further supports a hierarchical organization in which phonetic–articulatory information is most accessible before higher layers shift toward more abstract linguistic representations. 

These findings suggest that articulatory constraints are recoverable from acoustics through large-scale representation learning, opening practical avenues for applications in low-resource articulatory modeling, clinically oriented speech analysis and pronunciation assessment of L2 learners. Future work will investigate fine-tuning multilingual encoders on the target languages, combining representations across models, and employing nonlinear or sequential probing methods to better capture gestural timing.

\section{Acknowledgments}
This research has been initially conducted at the University of Eastern Finland and received funding from the BPI project THERADIA at a later stage.




\section{Generative AI Use Disclosure}
During the preparation of this work, the authors used DeepL and ChatGPT to assist with the editing of the English in some paragraphs. After using these tools, the authors reviewed and edited the content and take full responsibility for the publication.





\bibliographystyle{IEEEtran}
\bibliography{mybib}

\end{document}